# Critical Exponents of the Classical Heisenberg Ferromagnet

In a recent letter, Brown and Ciftan (BC) [1] reported high precision Monte Carlo (MC) estimates of the static critical exponents of the classical 3D Heisenberg model. While their finite-size scaling (FSS) analysis yields values for the critical temperature $T_c = 1/K_c$ and the critical exponent ratios $\beta/\nu, \gamma/\nu$, which are compatible with all recent findings, BC claim that the specific heat $C$ of this model is divergent at $T_c$, which is in strong disagreement with other recent high statistics MC simulations, high-temperature series analyses, field theoretic methods [2–5], and experimental studies [6], which all find a finite cusp-like behavior.

In their Ansatz (2) BC use a non-linear six parameter fit to 14 data points for $C$ on lattices of linear size $2 \leq L \leq 32$. The fit resulting in $\alpha/\nu = 0.117(4)$ has still a total $\chi^2 \approx 60$, and therefore is by standard reasoning not acceptable. This indicates that either the statistical errors of their data are underestimated or the use of such small lattices as $L = 2$ requires the inclusion of even more correction terms. Alternatively one may ask how the fit parameters would change if the smallest lattices are successively discarded. By hyperscaling BC deduce from this value a non-standard exponent $\nu = 0.642(2)$, leading in turn to non-standard estimates of $\beta$ and $\gamma$.

We find it very dangerous to base such an incisive conclusion solely on the very delicate FSS behavior of the specific heat. In particular we strongly disagree with the statement of BC that $\nu$ is extremely difficult to measure directly. The derivative of the Binder parameter $dU/dK$, and the logarithmic derivatives $d\ln\langle m\rangle/dK$ and $d\ln\langle m^2\rangle/dK$ all scale like $L^{1/\nu}$ and, using fluctuation formulas, can be as easily measured as $C$, and the statistical errors are straightforward to control. Already our data for $dU/dK$ in Ref. [2] gave an estimate of $\nu = 0.704(6)$ which agrees with our estimate of $\alpha/\nu$ by hyperscaling. Moreover it is compatible with the value $\nu = 0.698(2)$ derived from fits to the critical behavior of independent correlation length data in the high-temperature phase [2]. In Ref. [4] we studied this model with emphasis on topological excitations on much larger lattices with $8 \leq L \leq 80$. By analyzing the new data with the above three quantities at $K = 0.6930 \approx K_c$ we obtain from FSS fits a prediction of $\nu = 0.699(3)$ (cp. Fig. 1).

For our large lattices of Ref. [4] a fit of the cusp form $C(L) = C^{\rm reg} + C_0 L^{\alpha/\nu}$ to 12 data points yielded $\alpha/\nu = -0.225(80)$ ($\chi^2 = 7.84$). A more precise estimate can be obtained by analyzing the energy directly according to $E(L) = E^{\rm reg} + E_0 L^{(\alpha-1)/\nu}$, which definitely has a regular term. This yielded again a negative exponent ratio $\alpha/\nu = -0.166(31)$ ($\chi^2 = 11.3$), which by hyperscaling is completely consistent with our value of $\nu$. If we allow for a confluent correction $E_1 L^{(\alpha_1-1)/\nu}$, then the fit only slightly improves ($\chi^2 = 10.9$) with an almost vanishing amplitude $E_1 = -0.8 \times 10^{-13}$, showing that there really is no need for an additional term. If we add a term $C_1 L^{\alpha_1/\nu}$ to the $C$-fit, again we only get a marginally improved fit ($\chi^2 = 7.79$), but now with a positive exponent, $\alpha/\nu = 0.09(1)$, and $\alpha_1/\nu = -0.6(1)$. However, there are also different solutions with about the same $\chi^2$, showing the danger of being misled by a too flexible $C$-fit Ansatz.

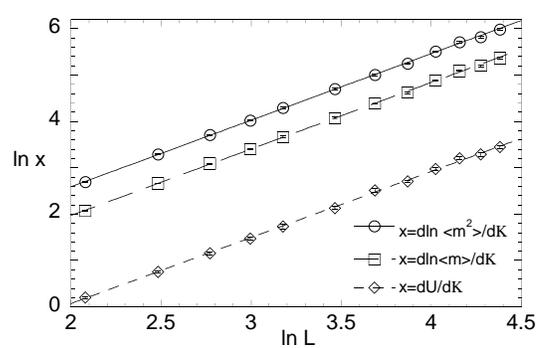

FIG. 1. Double logarithmic plots of data and fits at $K = 0.6930$ for $dU/dK$ (yielding $\nu = 0.703(6)$), $d\ln\langle m\rangle/dK$ ($\nu = 0.698(3)$), and $d\ln\langle m^2\rangle/dK$ ($\nu = 0.697(3)$) versus $L$.

The "universality scaling" of Fig. 3 in [1] indeed looks best for $\nu = 0.642$, but this plot cannot serve as an independent determination of $\nu$. The predicted data collapse is an *asymptotic* statement for large $L$ near $T_c$, and neither should it be expected far beyond $T_c$, nor for very small lattice sizes, even if the correct exponents are used.

The numerical coincidence that $K_c \approx \ln(2)$ should be interpreted with extreme care and by no means implies that the model is exactly solvable. Even for the simpler 3D Ising model no analytical solution is known – and the 6th digit of a similar conjecture for $K_c$ has recently been questioned numerically [7].

To summarize, we see no compelling reason that one should have any doubts that the estimates of the critical exponents of the 3D Heisenberg model, as consistently obtained by field theory, high-temperature series expansions, and MC methods, need to be reconsidered.

Christian Holm
  Institut für Theoretische Physik, FU Berlin
  Arnimallee 14, 14195 Berlin, Germany

Wolfhard Janke
  Institut für Physik, Johannes Gutenberg-Universität
  Staudinger Weg 7, 55099 Mainz, Germany